\documentclass[conference]{IEEEtran}
\IEEEoverridecommandlockouts
\usepackage{cite}
\usepackage{balance}
\usepackage{soul}
\usepackage{amsmath,amssymb,amsfonts}
\usepackage{algorithmic}
\usepackage{graphicx}
\usepackage{textcomp}
\usepackage{xcolor}
\usepackage{tikz}
\usepackage{booktabs} 
\def\BibTeX{{\rm B\kern-.05em{\sc i\kern-.025em b}\kern-.08em
    T\kern-.1667em\lower.7ex\hbox{E}\kern-.125emX}}
\begin{document}

\title{On the Physical Plausibility and Distribution Alignment for Sim-to-Real RF Positioning}

\author{
\IEEEauthorblockN{ 
Ararat Saribekyan\IEEEauthorrefmark{1}\IEEEauthorrefmark{2}\IEEEauthorrefmark{4},
Armen Manukyan\IEEEauthorrefmark{1}\IEEEauthorrefmark{2},  
Hrant Khachatrian\IEEEauthorrefmark{1}\IEEEauthorrefmark{2},
Theofanis P. Raptis\IEEEauthorrefmark{3}
}
\IEEEauthorblockA{
\IEEEauthorblockA{\IEEEauthorrefmark{1}Yerevan State University, Yerevan, Armenia.}
\IEEEauthorblockA{\IEEEauthorrefmark{2}YerevaNN, Yerevan, Armenia. Email: \{ararat, armen, hrant\}@yerevann.com}
\IEEEauthorblockA{\IEEEauthorrefmark{4}American University of Armenia, Armenia.}
\IEEEauthorrefmark{3}Institute of Informatics and Telematics, National Research Council, Pisa, Italy. Email: theofanis.raptis@iit.cnr.it
}
}

\maketitle
\begin{tikzpicture}[remember picture,overlay]
\node[anchor=south,yshift=10pt] at (current page.south) {\fbox{\parbox{\dimexpr\textwidth-\fboxsep-\fboxrule\relax}{
  \footnotesize{
     \copyright 2026 IEEE. Personal use of this material is permitted.  Permission from IEEE must be obtained for all other uses, in any current or future media, including reprinting/republishing this material for advertising or promotional purposes, creating new collective works, for resale or redistribution to servers or lists, or reuse of any copyrighted component of this work in other works.
  }
}}};
\end{tikzpicture}

\begin{abstract}
Reliable radio frequency (RF) positioning from cellular measurements is limited by the high cost and limited coverage of real drive-test data, especially when models must work on streets not seen during training. Previous work showed that ray tracing simulations can provide useful synthetic data for pretraining deep positioning models. In this paper, we focus on the simulation side and study how base-station calibration, physical realism, synthetic-data scale, and RSSI distribution alignment affect transfer to real data. Using a Sionna reconstruction of a Rome deployment, we calibrate each base station by adjusting its location, height, azimuth, and transmit power. We compare physically plausible calibrations with unconstrained ones that allow unrealistic base-station placements. We also compare deployment-specific synthetic data with much larger city-scale datasets. Although unconstrained calibration matches measured RSSI better, it does not always improve positioning accuracy. All synthetic pretraining approaches improve performance on known streets, with the best result obtained using city-scale unconstrained data. However, larger synthetic datasets alone do not improve performance on unseen streets. The best results on held-out streets are achieved only after normalizing simulated RSSI values to better match the real distribution. Overall, the results suggest that distribution alignment is more important than physical realism or dataset size for sim-to-real RF positioning.
\end{abstract}


\section{Introduction}
RF positioning is attractive for outdoor localization because radio fingerprints can be available in many deployed cellular and wireless networks, including settings where GPS or vision may be unreliable or expensive \cite{8970312}.
In fingerprint-based positioning, a model receives signal measurements from nearby base stations and predicts the receiver position. The practical obstacle is that real drive-test measurements are sparse, costly to collect, and tied to a particular deployment. For example, in the Rome dataset studied in~\cite{ali2022large,denardis2023outdoor}, a real-only positioning model can still lead to errors even on the known-street split, and generalization to held-out streets is substantially harder.

Synthetic radio propagation offers a natural way to increase supervision without collecting new drive-test data \cite{9354041}. Prior work~\cite{manukyan2025bridging} on this Rome dataset introduced the A/B/B'/C dataset hierarchy, the MapRadioFormer+ positioning backbone, Sionna RT simulation, Gaussian-process base-station calibration, and large-scale synthetic pretraining for sim-to-real RF positioning. We build on that setting by focusing on the simulation pipeline itself: How should the base stations be calibrated, what is the importance of physical plausibility, when does scale help, and how much does RSSI distribution alignment matter?

In~\cite{manukyan2025bridging}, the synthetic dataset optimized base-station position, height, and azimuth to improve rank correlation with measured RSSI. In this work, we extend that calibration by adding transmit power, which should not be constant across antennas, optimizing a log-domain RSSI RMSE objective. We compare two regimes:

\begin{itemize}
\item Constrained: base-station locations, orientations, and powers are optimized under physical plausibility constraints, such as keeping base stations on buildings and avoiding implausible heights.
\item Unconstrained: base-station locations, orientations, and powers are optimized without those constraints, allowing physically implausible but potentially more task-aligned effective simulator parameters.
\end{itemize}

The unconstrained simulator matches the measured RSSI more closely, but it does so by using effective base-station parameters that need not correspond to a physically realizable deployment. The constrained simulator is more plausible, but it pays an RSSI-fitting cost. The downstream question is therefore empirical: does physical plausibility improve real positioning transfer, or is distributional alignment between synthetic and real RSSI more important?

We also revisit the scale-versus-quality question studied in prior work~\cite{manukyan2025bridging}. We compare deployment-specific synthetic data, which remains close to the measured Rome geography and base-station context, with much larger city-scale synthetic data generated from calibrated base-station priors. This lets us test whether scaling synthetic supervision beyond the measured deployment improves real transfer, or whether additional synthetic data is useful only when its RSSI distribution is aligned with the real one.

Our results provide three insights. First, unconstrained calibration better matches measured RSSI, but the physically implausible fit does not consistently transfer better. Second, all synthetic-pretraining variants significantly improve known-street positioning over real-only training. Third, raw city-scale synthetic data does not produce a significant held-out-street improvement; the lowest held-out-street error appears only after normalizing the simulated RSSI values to better match the real distribution. These findings suggest that physical plausibility and synthetic scale are insufficient on their own, and that distribution alignment is a critical factor in sim-to-real RF positioning.

\section{Related Work}

RF-based positioning can be formulated as a fingerprinting problem, where received signal measurements from multiple transmitters are used as location-dependent signatures. Early systems such as RADAR and Horus showed that received signal strength measurements can be used for practical positioning by comparing online observations with a previously constructed radio map \cite{bahl2000radar,youssef2005horus}. Subsequent surveys have discussed a broad range of wireless positioning techniques and highlighted the importance of fingerprint density, environmental stability and measurement quality for positioning accuracy \cite{liu2007survey,he2016wifi,khalajmehrabadi2017modern}.

Learning-based positioning methods replace explicit fingerprint matching with models that learn the mapping from radio measurements to receiver coordinates. Deep learning has been applied to RSSI- and CSI-based fingerprints, with early works showing that neural models can extract useful representations from high-dimensional wireless measurements \cite{wang2016deepfi}. More recent work has studied different input representations for wireless positioning, including path-based features, and image-like spatial representations, showing that the structure of the RF input strongly affects the positioning model \cite{darbinyan2023mlnlos}. In this paper, we use the positioning architecture and training objective introduced in \cite{manukyan2025bridging} as the common backbone for all experiments. Our contribution is therefore not a new positioning architecture, but an analysis of how different synthetic pretraining distributions affect the same downstream positioning pipeline.

Synthetic wireless data generation has become an important tool for reducing the cost and sparsity of real measurement collection. Ray-tracing simulators can generate dense radio measurements over large geographic areas by combining scene geometry, antenna configuration, and propagation modeling. Sionna RT provides a differentiable and GPU-accelerated ray-tracing framework for radio propagation simulation, enabling the generation of channel responses and radio maps under controllable transmitter, receiver, antenna, and scene parameters \cite{hoydis2023sionnart}. Synthetic datasets like WAIR-D~\cite{waird} have also been used in learning-based RF tasks such as NLOS positioning and radio-map estimation, where simulated data provides scalable supervision that would be expensive to collect in the real world \cite{darbinyan2023mlnlos}.

However, the usefulness of ray-traced data depends on how well the simulator represents the real deployment. In practice, errors in map geometry, material properties, antenna positions, antenna orientations, transmit powers, and propagation assumptions can introduce a substantial mismatch between simulated and measured RF fingerprints. Recent work on differentiable ray tracing has shown that simulation parameters such as materials, scattering behavior, and antenna-related quantities can be calibrated using measurements \cite{hoydis2023sionnart}. At the same time, studies of outdoor urban RF simulation indicate that even careful ray-tracing configurations may remain insufficient for fully matching real measurements, and that base-station placement and antenna configuration can have a strong influence on downstream RF tasks \cite{manukyan2025limitations}.

This paper focuses on the role of such calibration choices in synthetic pretraining for RF positioning. Most prior works use either measured fingerprints or a single synthetic generator. In contrast, we compare two calibrated simulation regimes that differ in their treatment of physical plausibility. Furthermore, we show that validation-set choice and input-signal distribution alignment can substantially affect sim-to-real transfer.

\section{Methodology}

Our study is organized around a single principle: we hold the positioning
model, its training objective, and the fine-tuning and evaluation protocol fixed
across all conditions, and vary only the synthetic pretraining data. Any
difference in transfer is therefore attributable to the data-generation choices
rather than to architecture or optimization. The pipeline has three decoupled
stages: base-station calibration, synthetic data generation, and positioning
transfer.

We vary three factors, each isolated as far as the data permit:
\begin{itemize}
\item \emph{Physical plausibility} of the calibrated transmitters, via a
\emph{constrained} regime that enforces rooftop placement and realistic
heights and an \emph{unconstrained} regime that minimizes signal error alone
(Subsection~\ref{subsec:calibration}).
\item \emph{Deployment scope}, contrasting deployment-specific synthetic data
that reuses the real base stations and geometry of the measured network
(Dataset~B) with city-scale data on unseen geometry whose transmitters are
drawn from the calibrated priors (Dataset~C). This contrast tests whether
calibration knowledge \emph{transfers} beyond the measured deployment, rather
than scale in isolation, since the two sources differ in geometry and
base-station placement as well as in size.
\item \emph{Distribution alignment} between synthetic and real measurements,
probed by an input-normalization variant that standardizes the simulated
signals with their own statistics instead of the real-domain statistics used
elsewhere (Subsection~\ref{subsec:c_norm}).
\end{itemize}

\paragraph*{Calibration}
We treat the real Rome drive-test corpus, Dataset~A, as a calibration target.
For each real base station we optimize a horizontal offset, antenna height,
azimuth, and transmit power so that simulated RSSI matches the measured RSSI at
served UE locations, yielding two calibrated parameter sets, one per regime.
The full parameterization, objective, and optimizer are given in
Subsection~\ref{subsec:calibration}; the recovered marginal distributions over
height, power, and azimuth serve as the priors for city-scale generation.

\paragraph*{Generation}
The two regimes drive two kinds of synthetic supervision. Dataset~B deploys
the calibrated configuration exactly at the measured sites, keeping the real
base stations and scene geometry of Dataset~A and replacing only the
signals, and then deploys the UEs, which are sampled synthetically along the street network.
Dataset~C is fully synthetic at city scale: both base stations and UEs are
placed on previously unseen geometry, with transmitter parameters resampled from
the calibrated priors of each regime.

\paragraph*{Transfer and evaluation}
Every condition uses the same backbone~\cite{manukyan2025bridging}: we pretrain
on a synthetic source (B or C, optionally with the normalization variant)
and fine-tune on real Dataset~A. We evaluate on real known-street and
held-out-street splits, where the held-out split is constructed by removing any
training crop that overlaps the held-out region, making it a strict test of
geographic generalization. All reported metrics for a configuration come
from the last checkpoint of the best hyperparameter. We test two methods of choosing the best hyperparameter: one favoring known-street and one favoring held-out-street validation.

\section{The Dataset}
\label{sec:dataset}

We use three dataset families. Dataset~A, taken from~\cite{ali2022large,denardis2023outdoor}, is the real measured deployment and is used for calibration, fine-tuning, validation, and testing. Dataset~B is a deployment-specific synthetic dataset that uses the real BS locations and scene geometry of Dataset~A with synthetic street-sampled UEs. Dataset~C contains fully synthetic city-scale scenes with synthetic base stations and street-sampled UEs.
The generation procedures for Datasets B and C follow the prior work~\cite{manukyan2025bridging} loosely, but differ in calibration, UE filtering, and constrained/unconstrained simulation regimes.

\subsection{Base-Station Calibration}
\label{subsec:calibration}

Dataset~A anchors this work: a Rome drive-test corpus of roughly two thousand UE
locations served by the real cellular network, with four RF indicators (RSSI,
NSINR, NRSRP, NRSRQ) recorded per BS--UE pair. We use
it as a calibration target, tuning each synthetic base station's free antenna
parameters so that simulated propagation reproduces the measured signal.
Calibration runs under two regimes that differ only in whether a base station is
held to physically plausible rooftop placement: a \emph{constrained} and an
\emph{unconstrained} regimes. Plausibility and downstream utility are not the same objective:
relaxing placement realism lets the unconstrained regime absorb error sources the
simulator does not model (e.g. propagation physics, OSM-geometry inaccuracy, material
uncertainty, antenna-metadata error) into the recovered parameters, whereas the
constrained regime forgoes some of that freedom to keep deployments geometrically
realistic.

\paragraph*{Parameterization}
For each base station $i$ we optimize a five-dimensional vector
\begin{equation}
  \theta_i = (\Delta x_i,\ \Delta y_i,\ h_i,\ \phi_i,\ P_i),
\end{equation}
a horizontal offset $(\Delta x_i,\Delta y_i)$ from the surveyed position,
antenna altitude $h_i$, boresight azimuth $\phi_i$, and transmit power $P_i$.
The search box is $\Delta x_i,\Delta y_i \in [-100,100]$~m,
$h_i \in [15,40]$~m, $\phi_i \in [0,360)^{\circ}$, and
$P_i \in [20,50]$~dBm. Propagation parameters are held fixed during calibration:
a carrier of $1.2$~GHz, a $6{\times}6$ TR-38.901 transmit panel against a
cross-polarized dipole receiver, and up to $10^4$ paths per source. Relative to
the calibration of prior work~\cite{manukyan2025bridging}, which searched over
lateral offset, height, and azimuth alone, this formulation adds transmit power
as a fifth free parameter, narrows the height range to a physically motivated
$[15,40]$~m, and replaces the rank-correlation objective with a log-domain
fidelity term defined below.

\paragraph*{Fidelity term}
For a candidate $\theta_i$ we ray-trace at every served UE and compare predicted
to logged RSSI in the log domain,
\begin{equation}
  \mathrm{RMSE}(\theta_i) =
  \sqrt{\frac{1}{|\mathcal{U}_i|}
  \sum_{u \in \mathcal{U}_i}
  \big(\widehat{\mathrm{RSSI}}_u(\theta_i) - \mathrm{RSSI}_u\big)^2},
\end{equation}
over the valid UE set $\mathcal{U}_i$ (links with missing ground truth are dropped;
no floor or clipping is applied, so the two signals are compared directly).
Spearman rank correlation is logged for diagnostics but never optimized (in contrast to \cite{manukyan2025bridging}).

\paragraph*{Two regimes}
The unconstrained regime minimizes the data term alone,
\begin{equation}
  \mathcal{L}^{\mathrm{unc}}_i = \mathrm{RMSE}(\theta_i),
\end{equation}
leaving the antenna free in position, height, and power. The constrained regime
augments the objective with two physical-realism penalties,
\begin{equation}
  \mathcal{L}^{\mathrm{con}}_i =
  \mathrm{RMSE}(\theta_i) + P_{xy}(\theta_i) + P_{\mathrm{roof}}(\theta_i),
\end{equation}
where $P_{xy}(\theta_i) = 100$ when the optimized horizontal position does not
project onto a building footprint (tested by casting a downward ray against the
scene mesh) and $0$ otherwise. The rooftop term, with detected roof height
$z_r$ and a $10$~m mast margin,
\begin{align}
\Delta_{\mathrm{low}} &= \max(0, z_r-h_i), \\
\Delta_{\mathrm{high}} &= \max(0, h_i-(z_r+10)),
\end{align}

\begin{equation}
P_{\mathrm{roof}}(\theta_i)
=
\min\left\{
100,\;
\frac{1}{2}
\left(
\Delta_{\mathrm{low}}^2
+
\Delta_{\mathrm{high}}^2
\right)
\right\}.
\end{equation}
penalizes the antenna for falling below the roof or rising more than $10$~m
above it, with a flat dead-band in between and a cap at $100$. Both penalties are
soft, so a sufficiently large RMSE reduction can override either of them.

\paragraph*{Optimizer}
Each objective evaluation requires a full ray-tracing pass over the served UEs, so
we minimize it with Gaussian-process Bayesian optimization (expected-improvement
acquisition, $15$ space-filling initial points, $100$ evaluations, fixed seed),
independently for each of the $166$ base stations per regime. One caveat is unique
to this stage: to save compute, the optimizer uses a lighter solver than
generation: path depth $3$ with specular reflection disabled, versus depth $5$
with specular reflection enabled for Datasets~B and~C, so antennas are
calibrated under somewhat lighter propagation than they are later deployed under.

\paragraph*{Calibration results}
Mean optimized per-BS RMSE is $42.5$~dB unconstrained versus $44.2$~dB
constrained, a $1.7$~dB difference for physical plausibility. Table~\ref{tab:calib_dist}
shows where the regimes diverge. Altitude is the clearest effect: unconstrained
antennas pile up at the $40$~m ceiling ($61\%$), while the rooftop penalty pulls
constrained ones to realistic heights (median $34.8$~m). Power and azimuth behave
alike across regimes: power spans the full range and azimuth is effectively
uniform, since the symmetric array and weak per-link signal leave boresight
unidentified. Offsets saturate near the search-box boundary in both, so the
horizontal degrees of freedom absorb scene-and-survey mismatch rather than
recovering true positions.

Unfortunately, the fits are loose. A
typical per-BS systematic offset of $\sim\!31$~dB between simulated and measured
RSSI reflects a strong structural mismatch between Sionna RT and the real Rome
network. The recovered parameters are thus effective, not physical.

\begin{table}[t]
  \centering
  \caption{Recovered per-BS parameter distributions in the two calibration
  regimes. Ceiling = $40m$ }
  \label{tab:calib_dist}
  \small
  \begin{tabular}{@{}lll@{}}
    \toprule
    Parameter & Unconstrained  & Constrained  \\
    \midrule
    Altitude (m) &
      \begin{tabular}[t]{@{}l@{}}mean $36.0$, med.\ $40.0$\\$61\%$  ceiling\end{tabular} &
      \begin{tabular}[t]{@{}l@{}}mean $31.8$, med.\ $34.8$\\$43\%$ ceiling, $10\%$ floor\end{tabular} \\
    \addlinespace
    Power (dBm) &
      \begin{tabular}[t]{@{}l@{}}mean $34.1$, med.\ $33.4$\\full $[20,50]$ span\end{tabular} &
      \begin{tabular}[t]{@{}l@{}}mean $36.7$, med.\ $38.0$\\full span\end{tabular} \\
    \addlinespace
    Azimuth ($^{\circ}$) &
      \begin{tabular}[t]{@{}l@{}}$\approx$\,uniform\\(mean $188$, std $119$)\end{tabular} &
      \begin{tabular}[t]{@{}l@{}}$\approx$\,uniform\\(mean $190$, std $115$)\end{tabular} \\
    \addlinespace
    Offset (m) &
      \begin{tabular}[t]{@{}l@{}}mean $116$, med.\ $124$\\\end{tabular} &
      \begin{tabular}[t]{@{}l@{}}mean $110$, med.\ $113$\end{tabular} \\
    \bottomrule
  \end{tabular}
\end{table}

The decisive observation for what follows is that the soft placement constraint
is satisfied at the optimum for only $34\%$ of constrained base stations; for
the remaining two-thirds the RMSE term overrides $P_{xy}$ and the antenna settles
off-building (the rooftop-height term, by contrast, is almost never active, with
$P_{\mathrm{roof}} \approx 0.06$ at the optimum on average). In other words, a
constraint expressed as a soft penalty is violated most of the time. This is
precisely what motivates the hard, construction-time rooftop placement adopted
for the fully synthetic generator of
Subsection~\ref{subsec:datasetC}, where on-roof placement is guaranteed
geometrically rather than encouraged by a penalty. From the per-BS solutions we
extract the regime-specific marginal distributions of altitude, transmit power,
and azimuth; these serve as the empirical priors for synthetic base-station
placement at scale.

\subsection{Dataset B: Synthetic Measurements at Real BS Locations}
\label{subsec:datasetB}

Dataset~B sits between the per-BS calibration of
Subsection~\ref{subsec:calibration} and the fully synthetic generator of
Subsection~\ref{subsec:datasetC}: it retains the real base stations and scene
geometry of the measured deployment and replaces only the signals. The base
stations of Dataset~A are grouped into their per-cluster scenes and instantiated
with the calibrated parameters recovered above (one variant per regime) so that
B deploys the calibration solution exactly, whereas Dataset~C only inherits
its marginal priors. The receivers, however, are synthetic: as in Dataset~C, we
sample UEs along the OSM road network within $250$~m of a base station, reject
points inside building footprints, and fix the receiver at $1.5$~m, pairing each UE with every base station in its cluster. To keep the regime comparison strictly paired, UEs are sampled once for both regimes. The scenes are rendered with the same depth-$5$, specular-plus-diffuse solver.

Relative to the calibrated synthetic corpus of prior
work~\cite{manukyan2025bridging}, B adds the constrained/unconstrained
contrast and applies stricter validity criteria: street-only UE sampling, removal of crops that would leak training-region geometry across a held-out border, and a requirement that each retained crop keep at least three base stations with valid RSSI. This
tightens the evaluation but shrinks the usable data, so we treat B as a
diagnostic anchor that stays close to the calibration rather than as our primary
synthetic source.

\subsection{Dataset C: City-Scale Synthetic Generation}
\label{subsec:datasetC}

Dataset~C tests whether the placement knowledge distilled from the calibration of
Subsection~\ref{subsec:calibration} transfers to large, fully synthetic
environments. Both base stations and UEs are synthetic, but base-station placement
follows the constrained and unconstrained priors above, making C a controlled
large-scale replication of the calibration study in two matched variants on
previously unseen geometry. It supersedes the smaller Dataset~C of the prior
work~\cite{manukyan2025bridging}.

\begin{figure}[t]
  \centering
  \includegraphics[width=\columnwidth]{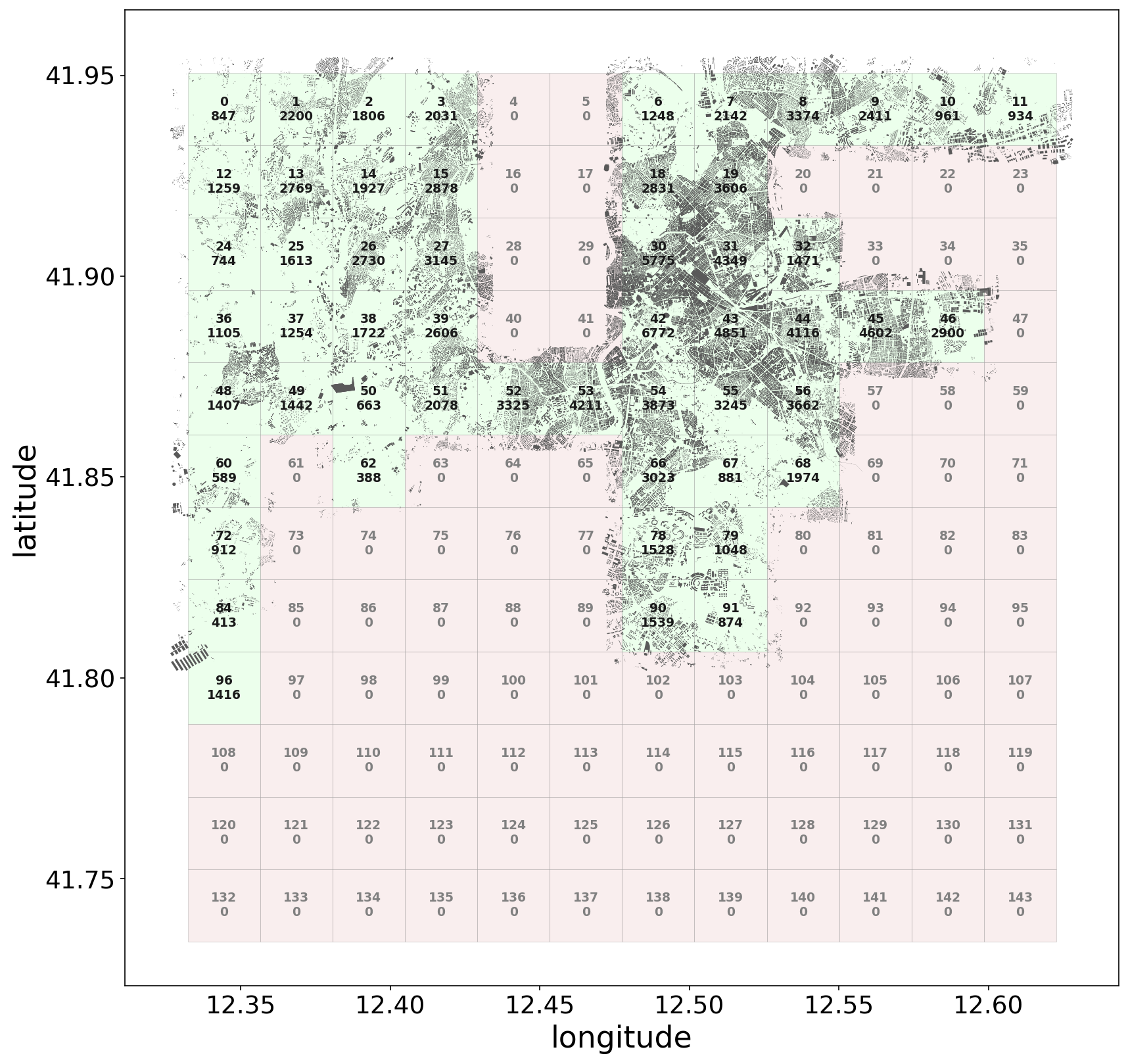}
  \caption{The $12 \times 12$ tiling of the
  $24\,\mathrm{km} \times 24\,\mathrm{km}$ bounding box over Rome. The
  $53$ patches retained for generation (those yielding a valid, non-empty
  building mesh) are highlighted; the remainder are river, parkland,
  periphery, or failed retrieval.}
  \label{fig:tiling}
\end{figure}

\paragraph*{Scenes Construction}
We take a $24\,\text{km} \times 24\,\text{km}$ square over Rome, centered at
$(41.8425^{\circ}\,\mathrm{N},\,12.4775^{\circ}\,\mathrm{E})$ and spanning latitude
$41.7344$--$41.9506^{\circ}$, longitude $12.3323$--$12.6227^{\circ}$, tiled into a
$12 \times 12$ grid of $144$ patches of about $2\,\text{km} \times 2\,\text{km}$.
Per patch, OSM building footprints are extruded to per-building heights into a triangulated mesh exported as a Sionna RT scene. The heights are taken from
\texttt{height} or \texttt{building:levels} tags where available, or use type-based
fallbacks. We keep only
patches where automated retrieval and extrusion yield non-empty, valid geometry:
$53$ of $144$ qualify, forming the contiguous urban core, while the rest are
dropped as river, parkland, periphery, or failed retrieval. Scenes are
content-addressed and cached, so both base-station variants reuse identical
geometry. Figure~\ref{fig:tiling} shows the tiling and the $53$ retained patches.

\begin{figure}[t]
  \centering
  \includegraphics[width=\columnwidth]{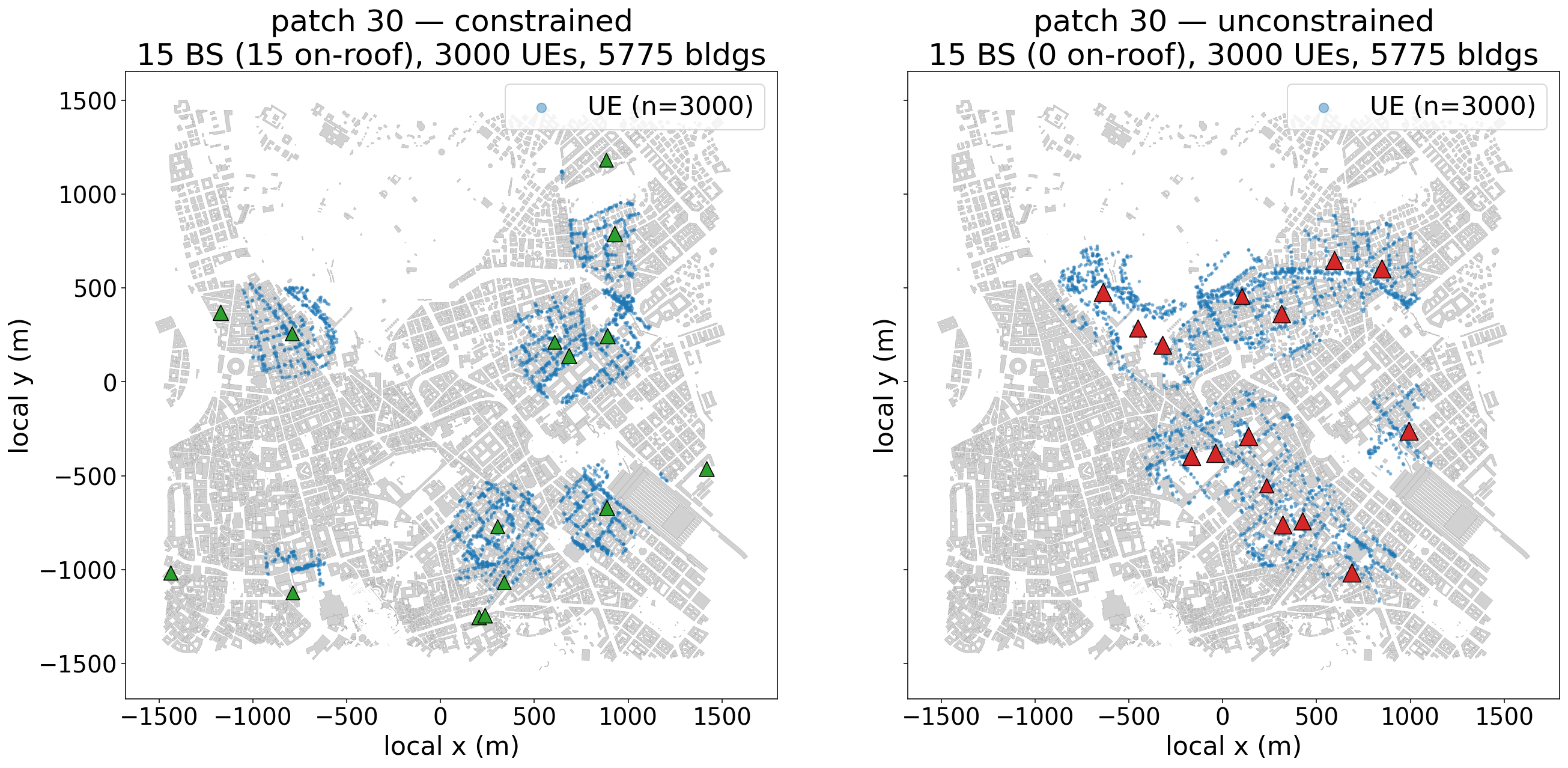}
  \caption{Base-station and UE placement on patch~$30$ under the constrained
  (rooftop) variant (left) and the unconstrained (free) variant (right). UEs are
  sampled independently per variant, against each variant's base-station placement.}
  \label{fig:patch30}
\end{figure}

\paragraph*{Synthetic base-station placement}
Each retained patch receives $15$ synthetic base stations whose parameters are
resampled from the calibration's empirical priors. These priors are the marginal distributions
over altitude, transmit power, and azimuth induced by the $166$ optimized base
stations in each regime of Subsection~\ref{subsec:calibration}. The two variants
realize the two regimes as hard placement rules rather than soft penalties. In the
\emph{constrained} (rooftop) variant, each base station is placed provably interior
to a building footprint, with altitude set to roof height plus a mast sampled in
$[3,6]$~m and clipped to the empirical band $[15,40]$~m---enforcing physical
plausibility by construction. In the \emph{unconstrained} (free) variant, the
horizontal position is free within the patch and altitude is drawn directly from
the unconstrained-regime distribution (mean $\approx 36$~m). Both variants draw
transmit power and boresight azimuth from the corresponding per-variant marginals.
The deployments thus reproduce each regime's measured statistics while occupying
geometry never seen during calibration. Figure~\ref{fig:patch30} contrasts the
variants on a representative patch.

\paragraph*{UE sampling}
We place $100\,000$ UEs per patch along the OSM road network, keeping only
positions within $250\,\mathrm{m}$ of at least one base station. We discard any point that falls inside a building
footprint and fix the receiver at a pedestrian height of $1.5\,\mathrm{m}$. Each
UE is paired with all $15$ base stations in its patch, giving $1.5\,\mathrm{M}$
(BS,~UE) pairs per patch. In contrast to B, at city scale we sample UEs independently for each variant, against that variant's base-station placement.

\paragraph*{Ray tracing and metrics}
Propagation is computed with Sionna RT\cite{hoydis2023sionnart} at a carrier of $1.2$~GHz, a maximum path
depth of $5$, and specular plus diffuse reflection, refraction, and
line-of-sight all enabled, using $10^6$ rays per source and at most $10^4$ paths
per source with the 3GPP TR~38.901 transmit pattern. Base stations are simulated
one transmitter at a time, so that each station's sampled transmit power is
applied exactly. For every pair we record RSSI, NSINR, NRSRP, and NRSRQ, with
RSSI floored at $-140$~dBm. Specular reflection at depth $5$ is necessary in the
dense urban canyons of central Rome: a shallower, diffuse-only solver leaves most
non-line-of-sight links without a valid path and spuriously floors roughly $90\%$
of the samples.

\paragraph*{Scale and statistics}
Each variant comprises about $53$ built patches $\times\,1.5\,\text{M}
= 79.5\,\text{M}$ (BS,~UE) pairs, roughly $159\,\text{M}$ pairs across both
variants.
Throughput is roughly $9$~s per $256$-UE batch and about $15$~h per
patch on an NVIDIA H100, parallelized across $24$ patch-range shards. The floored fraction ranges from about $45\%$ to
$99\%$ with building density: each UE is in range of only its nearest one or two
base stations yet is paired against all $15$, so the high floor is a property of
the sparse all-pairs deployment rather than a solver artifact.

\begin{figure*}
    \centering
    \includegraphics[width=0.99\linewidth]{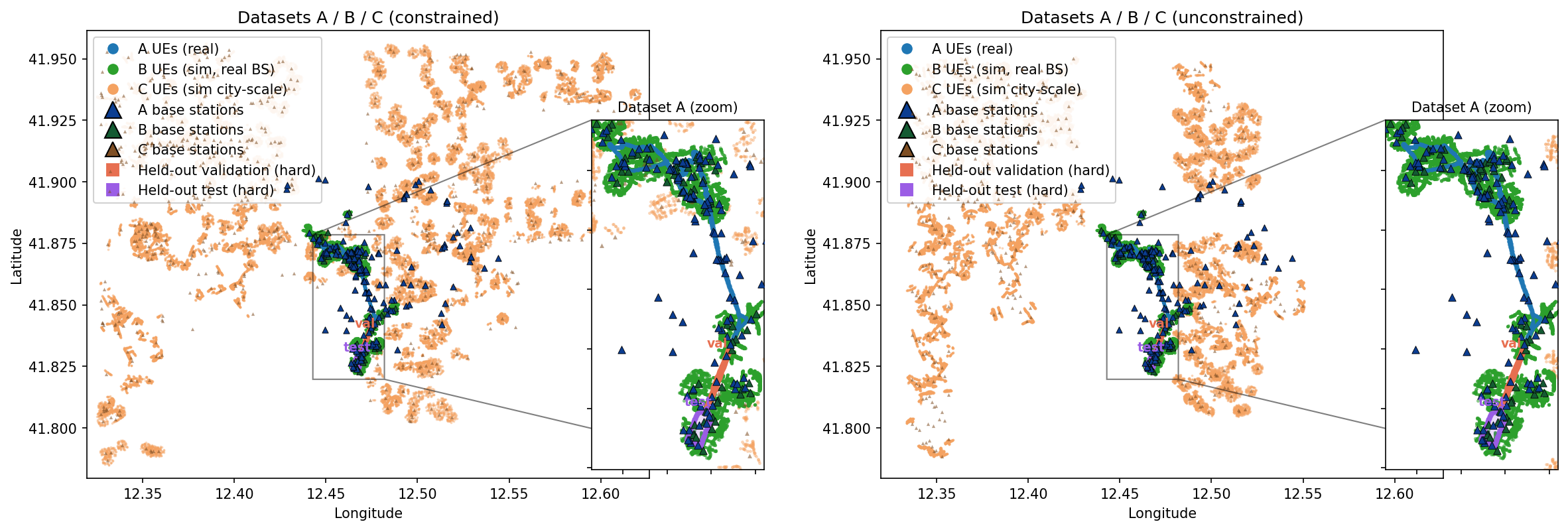}
    \caption{All datasets used in this work visualized on the map of Rome.}
    \label{fig:dataset}
\end{figure*}

All datasets in both regimes are visualized in Figure~\ref{fig:dataset}.

\subsection{C Constrained Normalization Variant} 
\label{subsec:c_norm}
\begin{figure}[t]
    \centering
    \includegraphics[width=\columnwidth]{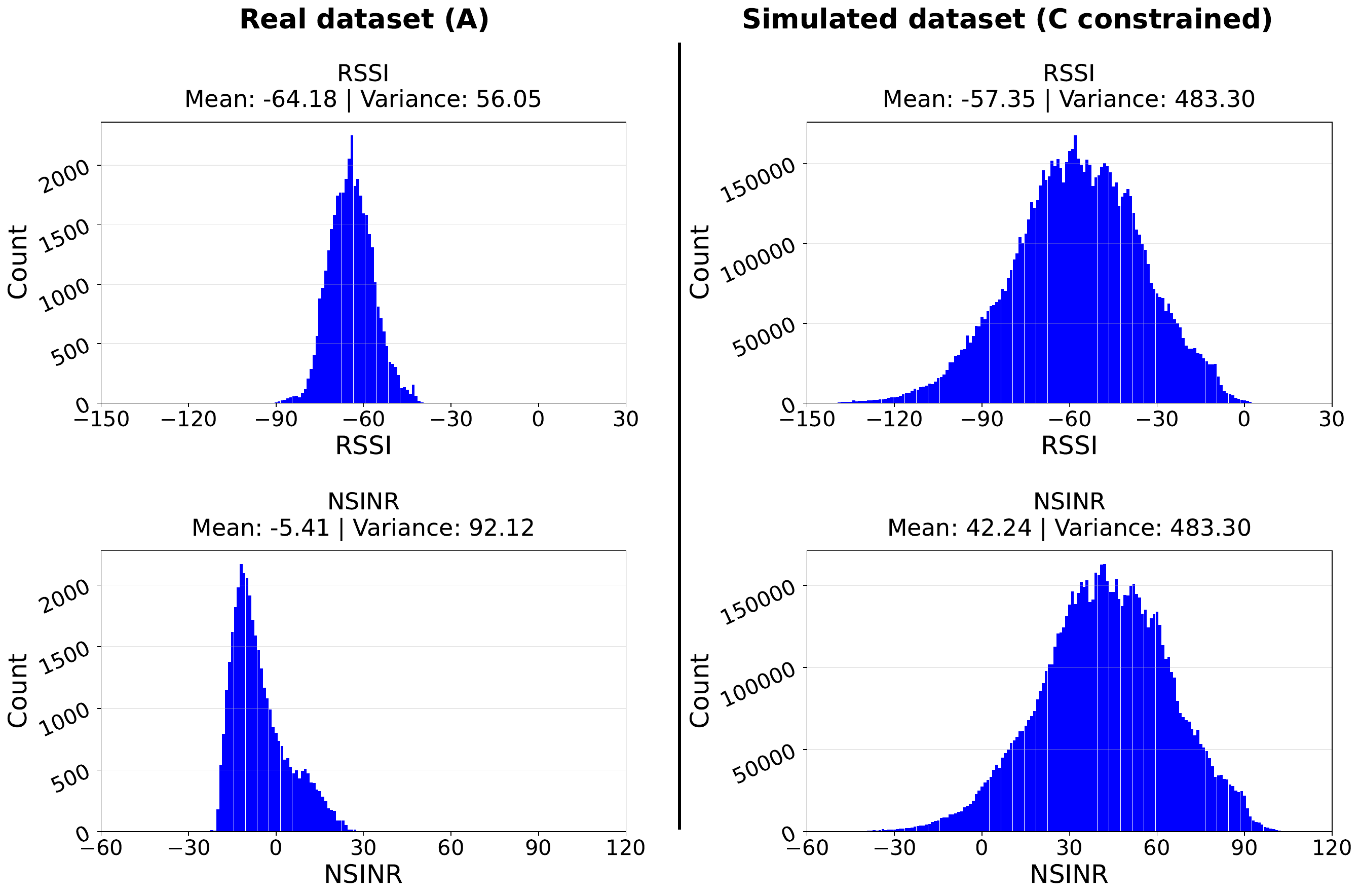}
    \caption{Measurement distribution comparison between real Dataset~A and the C-constrained synthetic dataset.}
    \label{fig:rssi_nsinr_distribution}
\end{figure}

In the main experiment, measurement values from both real and simulated datasets were normalized using the mean and variance estimated from the real Dataset A. This choice keeps the input scale tied to the target domain used for fine-tuning and evaluation. However, during the analysis of the generated data, we observed a mismatch between the measurement distributions of the real and simulated domains as illustrated in Fig~\ref{fig:rssi_nsinr_distribution}.

This observation motivated an additional ablation experiment for the C-constrained dataset. In this variant, instead of normalizing the simulated measurements using the statistics of Dataset A, we computed the mean and variance directly from the C-constrained dataset and used these statistics during synthetic pretraining. All other training and fine-tuning settings were kept unchanged. 

\section{Architecture and Training Details}

As a backbone, we use the model introduced in~\cite{manukyan2025bridging}. The architecture is based on a DINO-style feature encoder and is extended with additional task-specific blocks for processing RF fingerprint measurements and BS-related information. Given the measured signal values associated with the available transmitters, the network learns to map the observed radio-signal pattern to the receiver position, producing a two-dimensional coordinate estimate as output. 

For all training configurations, we use a common input preprocessing and optimization setup. The models are trained
using the Adam optimizer. From each dataset, we extract spatial crops of size $1000\times1000$ where one pixel corresponds to one meter, and resize them to $448\times448$ before feeding them to the network. Training is performed with a batch size of 4 per GPU on two NVIDIA H100 GPUs using Distributed Data Parallel training.

The training objective is the one used in~\cite{manukyan2025bridging}
\begin{equation}
-\frac{1}{|\Omega|}
\sum_{p \in \Omega}
\left[
Y_p \log(\hat{Y}_p)
+
(1-Y_p)\log(1-\hat{Y}_p)
\right]
\end{equation}
where $\Omega$ denotes the set of pixels in the output map, and $Y_p$ and $\hat{Y}_p$ are the target and predicted values at pixel $p$, respectively.

For pretraining on simulated data, we use the same optimization protocol for both $B$ and $C$. The learning rate follows a Warmup-Stable-Decay schedule~\cite{hu2024minicpmunveilingpotentialsmall}, with $3\%$ of the training steps used for warm-up, $10\%$ for decay. The stable learning rate is selected through a grid search based on validation loss. For dataset $B$, both the constrained and unconstrained variants are trained for $60$ epochs. For dataset $C$, the constrained and unconstrained variants are trained for a smaller number of epochs chosen so that, across the different simulated datasets and configurations, the model is exposed to a comparable amount of training data.

We evaluate six training conditions: real-only training from scratch, and five pretrainings followed by fine-tuning on Dataset A, namely: B constrained pretraining, B unconstrained pretraining, C constrained pretraining, C unconstrained pretraining, and C constrained pretraining with RSSI normalization. 

Fine-tuning on Dataset A follows the same input representation and positioning objective. We sweep over 18 hyperparameter configurations formed by six epoch budgets (10, 20, 40, 80, 100, 120) and three learning rates ($6\cdot10^{-5}$, $2\cdot 10^{-4}$, $6\cdot 10^{-4}$). All fine-tuning runs use cosine learning rate schedule with fixed 2-epoch warmups. Each run is evaluated from its final checkpoint on two splits: Known-Street Validation and Held-Out-Street Validation. 

As there is a significant distribution shift between known streets and held-out streets, we try two model-selection strategies for each condition. In the Known-Street Validation strategy, we choose the hyperparameter configuration with the lowest final-checkpoint Known-Street Validation error and report all validation and test metrics from that same final checkpoint. In the Held-Out-Street Validation strategy, we repeat the same procedure using Held-Out-Street Validation. 

The reported error is mean per-sample Euclidean positioning error in meters, computed between the argmax of the predicted heatmap and the true UE location. To quantify uncertainty, we report 95\% percentile bootstrap confidence intervals on test-set mean errors using 10,000 bootstrap resamples. For differences against the real-only A baseline, we use paired bootstrap resampling with identical sample indices across runs, because all runs are evaluated with batch size 1 and deterministic sample order. A confidence interval for a difference that excludes zero is treated as statistically significant.

\begin{table*}[t]
  \centering
  \caption{Positioning results selected by Known-Street Validation. Errors are in meters; lower is better. Test columns include 95\% bootstrap confidence intervals. Delta columns report paired differences against \textit{A}. * indicates significant difference.}
  \label{tab:medium_selection}
  \scriptsize
  \begin{tabular}{@{}lrrrrrr@{}}
    \toprule
    Setup & Held-Out Val & Known Val & Held-Out Test & Known Test & $\Delta$ Held-Out & $\Delta$ Known \\
    \midrule
    A (from scratch) & 386.96 & 125.08 & 296.15 [268.0, 325.0] & 123.80 [99.9, 151.3] & -- & -- \\
    B constrained $\rightarrow$ A & 410.18 & 62.00 & 219.23 [197.1, 243.0] & 60.18 [44.2, 80.6] & -76.91 [-106.7, -47.9]* & -63.62 [-82.9, -45.6]* \\
    B unconstrained $\rightarrow$ A & 379.85 & 65.53 & 242.47 [218.5, 267.1] & 42.04 [32.4, 52.8] & -53.67 [-82.2, -25.4]* & -81.76 [-108.9, -58.2]* \\
    C constrained $\rightarrow$ A & 420.23 & 56.69 & 289.89 [262.9, 318.2] & 51.21 [40.3, 63.6] & -6.26 [-32.4, +20.8] & -72.60 [-99.4, -49.8]* \\
    C unconstrained $\rightarrow$ A & 357.70 & 59.24 & 288.44 [257.6, 320.0] & \textbf{37.86 [29.1, 47.6]} & -7.71 [-41.6, +26.1] & -85.94 [-113.7, -62.1]* \\
    C constrained norm. $\rightarrow$ A & 397.40 & \textbf{55.29} & 282.48 [257.6, 307.5] & 55.98 [44.1, 69.3] & -13.67 [-47.0, +18.9] & -67.82 [-96.2, -42.9]* \\
    \bottomrule
  \end{tabular}
\end{table*}

\begin{table*}[t]
  \centering
  \caption{Positioning results selected by Held-Out-Street Validation. Errors are in meters; lower is better. Test columns include 95\% bootstrap confidence intervals. Delta columns report paired differences against \textit{A}. * indicates significant difference.}
  \label{tab:hard_selection}
  \scriptsize
  \begin{tabular}{@{}lrrrrrr@{}}
    \toprule
    Setup & Held-Out Val & Known Val & Held-Out Test & Known Test & $\Delta$ Held-Out & $\Delta$ Known \\
    \midrule
    A (from scratch) & 326.91 & 162.10 & 268.93 [246.4, 293.3] & 179.68 [154.6, 205.9] & -- & -- \\
    B constrained $\rightarrow$ A & 343.24 & 120.18 & 249.61 [225.4, 274.4] & 121.55 [97.6, 147.4] & -19.32 [-43.6, +4.5] & -58.13 [-80.4, -36.0]* \\
    B unconstrained $\rightarrow$ A & 329.76 & 101.17 & 275.37 [251.6, 300.3] & 89.29 [71.1, 109.6] & +6.44 [-23.1, +34.4] & -90.39 [-114.7, -65.9]* \\
    C constrained $\rightarrow$ A & 333.74 & 130.30 & 335.25 [311.3, 359.8] & 116.06 [98.1, 135.0] & +66.32 [+31.0, +101.1]* & -63.62 [-88.2, -39.0]* \\
    C unconstrained $\rightarrow$ A & 328.60 & 158.25 & 261.92 [241.0, 284.1] & 143.96 [120.1, 170.0] & -7.01 [-36.0, +21.5] & -35.72 [-58.6, -13.6]* \\
    C constrained norm. $\rightarrow$ A & 343.05 & 159.26 & \textbf{199.16 [178.3, 220.8]} & 158.20 [136.1, 181.9] & -69.76 [-98.9, -41.6]* & -21.48 [-43.8, -0.5]* \\
    \bottomrule
  \end{tabular}
\end{table*}

\section{Results}

\subsection{Base-Station Calibration}

Table~\ref{tab:calib_dist} summarizes the calibrated parameter distributions. The main finding is that unconstrained calibration gives a better RSSI fit, with mean best RMSE of 42.5~dB compared with 44.2~dB for constrained calibration. This 1.7~dB gap indicates that there is an inherent limitation in simulation fidelity that with physically plausible antenna locations we are unable to reproduce real measurements even to the degree of unrealistic setups. Also note that the large absolute errors indicate that the optimized base-station parameters should be interpreted as effective simulator parameters rather than recovered physical metadata.

\subsection{Transfer Under Known-Street Validation Selection}

Table~\ref{tab:medium_selection} reports the hyperparameter selected by Known-Street Validation error at the last checkpoint for each training condition. This selector favors longer training schedules (80+ epochs), as overfitting on the known streets does not harm the model's performance within the known area. Under this selection criterion, all synthetic-pretraining conditions substantially improve Known-Street Test over real-only training. The lowest Known-Street Test error is obtained by C unconstrained pretraining, reaching 37.86~m compared with 123.80~m with no pretraining.

While Known-Street Validation selection is misaligned with the objective of improving held-out performance, we report Held-Out-Street Test performance as well. Both B variants significantly outperform real-only training, but none of the city-scale C variants do so. Thus, scaling synthetic data beyond the measured deployment does not automatically improve held-out-street transfer.

\subsection{Transfer Under Held-Out-Street Validation Selection}

Table~\ref{tab:hard_selection} reports the same training conditions selected by the score on Held-Out-Street Validation at the last checkpoint. This selector favors much shorter synthetic training schedules, usually 10--20 epochs. The selected models again show significant gains on Known-Street Test. However, the Held-Out-Street Test conclusions change. Raw city-scale C pretraining, both constrained and unconstrained, does not significantly improve Held-Out-Street Test. Instead, pretraining on C constrained with RSSI normalization gives the lowest Held-Out-Street Test error, 199.16~m, a significant 69.76~m improvement over from-scratch training.

The difference between Tables~\ref{tab:medium_selection} and~\ref{tab:hard_selection} is informative. Known-Street Validation identifies B pretraining as a strong source, while Held-Out-Street Validation identifies normalized city-scale constrained simulation as the lowest-error held-out-street model. This suggests that validation design is an important part of sim-to-real RF positioning, and the geographic location where synthetic data is generated matters.

\subsection{Physical Plausibility Does Not Predict the Transfer}

Comparing constrained and unconstrained regimes sheds light on whether physical plausibility predicts downstream utility. There is a trade-off between fitting real-world RSSI signals and preserving physically plausible base-station locations. However, neither regime dominates in downstream positioning. The preferred regime varies across the model-selection criterion, the synthetic-data scale, and the evaluation split.

These results support the main qualitative claim of the paper: physical plausibility alone is not a reliable predictor of sim-to-real transfer.

\subsection{Distribution Alignment Through RSSI Normalization}

We include one RSSI-normalized experiment, C constrained normalized $\rightarrow$ A, to test whether reducing signal-scale mismatch between synthetic and real datasets can improve transfer. The result is informative: without normalization, C constrained $\rightarrow$ A significantly degrades Held-Out-Street Test under Held-Out-Street Validation selection; with normalization, the same constrained city-scale synthetic source yields the lowest Held-Out-Street Test error and significantly improves over real-only.

This suggests that distribution alignment can be more impactful than physical plausibility alone. Future work should evaluate normalization in the remaining training settings. This result should motivate further analysis on more aspects of sim-to-real gap.

\subsection{Relation to Prior Sim-to-Real Results}

Prior work on this dataset reported large gains from calibrated synthetic pretraining, including for held-out-street positioning~\cite{manukyan2025bridging}. Our results are more conservative. One likely reason is protocol strictness. \cite{manukyan2025bridging} performed the training vs. held-out split according to the location of the user equipment, which in border cases allowed overlap at the map-crop level. Here we explicitly filter out training samples that cover any portion of the held-out area, regardless of the device and antenna locations. Furthermore, we constrain UEs to valid street locations and retain only crops with sufficient base-station coverage. This setup makes the held-out-street split a more demanding test of geographic generalization.

On the other hand, this work sheds light on the potential impact of synthetic pretraining. It highlights new dimensions of sim-to-real gaps that can significantly impact the transfer quality.

\section{Conclusions}

We studied sim-to-real RF positioning through the lens of physical plausibility and distribution alignment. We calibrated Sionna RT base-station parameters under constrained and unconstrained regimes, generated deployment-specific and city-scale synthetic datasets, and evaluated transfer to real measurements with paired bootstrap confidence intervals. Unconstrained calibration improves raw RSSI RMSE, but downstream positioning shows that raw simulator fidelity and physical plausibility are insufficient predictors of transfer. The most consistent result is that synthetic pretraining improves the Known-Street Test split across all evaluated synthetic sources. Generalization to the held-out-street split is more limited, but there is a clear winner: the RSSI-normalized constrained city-scale pretraining is significantly more accurate than the real-only baseline. These findings suggest that future sim-to-real RF positioning work should treat physical calibration, synthetic scale, RSSI normalization, and validation split design as distinct factors and should look for methods to reduce the sim-to-real gap.

\section*{Acknowledgements}
This work was supported by funding under the bilateral agreement between CNR (Italy) and HESC MESCS RA (Armenia) as part of the DeepRF project for the 2025–2026 biennium, and by the HESC MESCS RA grant No. 22rl-052 (DISTAL).

\balance
\bibliographystyle{IEEEtran}
\bibliography{references}
\end{document}